\documentclass[trackchanges,longbib,twocolumn]{aastex701}

\usepackage{siunitx}

\usepackage[
    ]{starfont}
\usepackage{amsmath}

\DeclareSymbolFont{starfontsym}{OT1}{sts}{m}{n}
\DeclareMathSymbol{\mathSun}{\mathord}{starfontsym}{115}
\DeclareMathSymbol{\mathTerra}{\mathord}{starfontsym}{76}
\DeclareMathSymbol{\mathvarTerra}{\mathord}{starfontsym}{108}
\DeclareMathSymbol{\mathMoon}{\mathord}{starfontsym}{100}
\DeclareMathSymbol{\mathvarMoon}{\mathord}{starfontsym}{97}

\received{November 16, 2025}
\revised{March 17, 2026}
\accepted{March 19, 2026}

\submitjournal{PSJ}

\begin{document}

\title{Onset of habitable conditions on the Hadean Earth set by feedback between tides and greenhouse forcing}

\correspondingauthor{Marijn R. van Dijk}
\author[orcid=0009-0005-6677-1929, sname={van Dijk},gname=Marijn]{Marijn R. van Dijk}
\affiliation{Kapteyn Astronomical Institute, University of Groningen, Groningen, The Netherlands}
\email[show]{m.r.van.dijk.3@student.rug.nl}

\author[orcid=0000-0002-8368-4641, gname=Harrison, sname='Nicholls']{Harrison Nicholls}
\affiliation{Institute of Astronomy, University of Cambridge, Cambridge, United Kingdom}
\email{harrison.nicholls@ast.cam.ac.uk}

\author[orcid=0000-0002-3286-7683, gname=Tim,sname=Lichtenberg]{Tim Lichtenberg}
\affiliation{Kapteyn Astronomical Institute, University of Groningen, Groningen, The Netherlands}
\email{tim.lichtenberg@rug.nl}

\begin{abstract}

\noindent In the aftermath of the Moon-forming giant impact, the Hadean Earth's mantle and surface crystallized from a global magma ocean blanketed by a dense volatile-rich atmosphere. While prior studies have explored the thermal evolution of such early Earth scenarios under idealized, oxidizing conditions, the potential feedback between tidal heating driven by Earth–Moon orbital forcing and variable redox scenarios have not yet been explored in detail. We investigate whether tidal heating could have prolonged this early magma ocean phase and supported quasi-steady state epochs of global radiative equilibrium: periods of thermal balance between outgoing radiation and interior heat flux. Using the \texttt{PROTEUS} simulation framework, we simulate Earth's early evolution under a range of plausible tidal power densities, oxygen fugacities, and volatile inventories. Our results suggest that feedback between tidal heating and atmospheric forcing can induce substantial variation in magma ocean lifetimes, from $\sim$30\,Myr up to $\sim$500\,Myr, sensitive to interior redox conditions. Global radiative equilibrium epochs commonly arise across this range, lasting from $\sim$2 to $\sim$320\,Myr, and typically occur from 24\,Myr after the Moon-forming impact. Under oxidizing conditions, late-stage H$_2$O degassing promotes melt retention and sustained heating due to its significant contribution to greenhouse forcing. Weak tides increase the atmospheric abundance of H$_2$S and NH$_3$ and deplete CO. Therefore, the feedback between tides and atmospheric forcing induces a disequilibrium signature in the magma ocean atmosphere.

\end{abstract}

\section{Introduction} 

\noindent The leading hypotheses for Moon formation -- including accretion from a circumplanetary disk and giant impact scenarios -- predict a tremendous release of energy in the Earth, capable of inducing partial or complete (re)melting of the early Earth's interior \citep{Canup2020,NAKAJIMA2015286, Lock2018, ZAHNLE201574, Nakajima2021EPSL}. This process likely led to the formation of a global or regional magma oceans on both the Earth and the Moon, supported by compositional measurements of KREEP elements in lunar samples \citep{Borg2004,https://doi.org/10.1029/2000JE001336,https://doi.org/10.1029/RG017i001p00073}, the presence of distinct low-velocity zones deep in Earth's mantle \citep{labrosse_core_2003, willams_deep_1996, labrosse_deep_2007}, and Fe-isotope data suggesting a melted magma ocean cumulate component in the upper mantle \citep{Williams2021SciA}. The  thermal evolution of a primordial magma ocean is governed by thermal emission; in the cases of the Earth and Moon, it eventually solidified \citep{Nicholls_2024_JGRP, ABE199727, Hamano2013, Miyazaki2022}. 

Several studies have investigated the solidification of the Earth's early magma ocean by focusing heavily on its interior dynamics \citep{doi:https://doi.org/10.1029/GM074p0041,ABE199727,2000orem.book..323S,Monteux2016,Miyazaki2019JGRB}. Later work extended this by coupling atmospheric models to the interior, highlighting the significant role of atmospheric blanketing in delaying mantle cooling and affecting volatile evolution \citep{1985LPSC...15..545A, 1986EM&P...34..223M, ZAHNLE198862, Hamano2013, Nicholls_2024_JGRP, https://doi.org/10.1002/jgre.20068, 2010DPS....42.5205E, 2021AsBio..21.1325B, 2016ApJ...829...63S, KrissansenTotton2024, refId1, Lichtenberg2021, Miyazaki2022}. 

The evolution of the early atmosphere and mantle has implications for the prebiotic atmosphere and surface conditions of the early Earth, which set the chemical background environment for the origin of life and the onset of habitable conditions during the Hadean and Eoarchean \citep{2020CSysC...2...35B, Rout2025, 2002PNAS...9914628M, Sasselov2020SciA, Miyazaki2022, Ianeselli2022}. However, both the oxidation state of the mantle, and the redox conditions established during core formation after the giant impact, remain poorly constrained \citep{Lichtenberg2023PP7}. In particular, during and following the giant impact, the mantle, atmosphere, and core of the Earth and Moon are expected to equilibrate and undergo a series of redox steps \citep{Deguen2014,Landeau2016NatGeo,Landeau2021EPSL}, which are intricately linked to the degassing and composition of the primitive atmosphere \citep{Suer2023,Lichtenberg2025TOG3,Bower2025}.

Despite this uncertainty, studies of tidal heating, a process that influences both global heating rates and atmospheric composition, generally assume fixed atmospheric scenarios \citep[e.g.,][]{Rufu_2020, ZAHNLE201574}, leaving open questions about how tidal heating interacts with a coupled interior-atmosphere structure under different evolutionary scenarios. This is particularly important to consider during the Hadean, when the Moon was closer to the Earth and tidal effects would have been more pronounced \citep{ZAHNLE201574,farhat2025tides}. In this work, we incorporate a range of internal heating rates alongside our existing coupled interior-atmospheric evolution framework \citep{Lichtenberg2021,Nicholls_2024_JGRP}. We aim to better understand the interaction between tidal heating and climate evolution and their effects on the Earth's magma ocean evolution, solidification timescale and atmospheric composition. 

As the planet cools from an initially molten state, the melt fraction decreases, and the viscosity of the mantle increases \citep[e.g.,][]{costa2009model}. The increase in viscosity allows for shear forces to dissipate energy, so tidal heating becomes significantly more effective at melt fractions \citep{Hay_2019} near the critical melt fraction (between $0.2$ to $0.6$), around which the fluid dynamics transition from a liquid-like flow to viscous-creep \citep{doi:https://doi.org/10.1029/GM074p0041, SCOTT2006177, costa2009model}. The rate at which a planet radiates energy is dependent on several factors including stellar, orbital, and planetary properties. A strong greenhouse effect was likely contemporaneous with Earth's primordial magma ocean, set by an atmosphere of volatiles outgassed from the surface. Detailed simulations have shown that Hadean magma ocean solidification timescales can be substantially prolonged in the presence of such secondary atmospheres \citep{ABE199727, Hamano2013, hamano2015lifetime, Bower2022PSJ, Nicholls_2024_JGRP}.  Evolutionary models are thus necessary for modeling the crystallization sequence of early mantles, as physical dependency between melt-state and thermal evolution introduces an important hysteresis behavior \citep{ZAHNLE201574}. \citet{nicholls_tides_2025,Nicholls2025c} show that coupling between tidal forces and mantle redox state can greatly affect the longevity of magma ocean phases for the case of the L\,98-59 exoplanetary system. This motivates us in this study to consider renewed exploration of the Hadean magma ocean with a model of comparable physical and chemical detail. Here, we thus aim to expand upon this work through consideration of the Earth-Moon system. 

Lunar tides have dominated over solar tides throughout Earth's history \citep{Goldreich1966,Daher2021}. \citet{Heller2021} show that Earth could have been subjected to substantial tidal forces, as the Moon was closer and Earth's day-length was shorter \citep{ZAHNLE201574, SPALDING201928, Canup2020}. Lunar tides arise due to the Earth's faster rotation compared to the orbital motion of the Moon; the gravitational interaction between the bodies slows down Earth's rotation while exchanging angular momentum, causing an increase in the lunar orbital separation over time \citep{greenberg2009frequency,efroimsky2013tidal,1999ssd..book.....M}. Conceptually, the phase lag between lunar orbital motion and Earth's axial rotation increases for faster Earth rotation rates, so large amounts of `initial' terrestrial angular momentum $L_{\mathTerra}^\text{(ini)}$ will physically correspond to more tidal heating within the Earth.

The total angular momentum of the Earth--Moon system $L_{EM}$ need not have been constant over time \citep{Cuk2012,WISDOM2015138,Cuk2016,TIAN201790,Rufu_2020}. The frequent occurrence of approximately $4.35$\,Ga ages among lunar surface samples and a spike in lunar zircon ages at about the same time are indicative of a remelting event driven by the Moon’s orbital evolution, supportive of angular momentum dumping onto the Sun \citep{Nimmo2024, doi:10.1126/sciadv.adn9871}. Moreover, recent developments in Moon formation scenarios favor cases with high initial terrestrial angular momentum as they help to explain similarities between the Earth's and Moon's compositions \citep{Canup2020, https://doi.org/10.1002/2016JE005239, Lock2018}. \citet{Lock2018} introduce the so-called Synestia scenario, based on a lower bound of the total initial angular momentum of $L_{EM}^\text{(ini)} > 1.7$ times the current value of the Earth--Moon system found by \citet{kokubo2010formation}. Given this, the initial angular momentum of the Earth--Moon system may have been larger than its current value, allowing for more tidal heating to occur within the young Earth and Moon.  

Secondary atmospheres limit heat loss and thereby prolong large melt fractions arising from formation \citep{elkins2008linked,salvador2023magma}, so tidal dissipation may have been inefficient during the early stages of the solidification. As such, substantial greenhouse effects would reduce tidal dissipation efficiency, resulting in smaller angular momentum exchange rates between the Earth and Moon, and yield a reduced lunar orbital recession rate \citep{ZAHNLE201574}. Initial orbital separations may have been able to maintain relatively fixed tidal power densities within the Earth. As such, here we consider a range of  constant tidal power densities, simulated independently, to quantify the sensitivity of lunar orbital recession to physical and chemical processes on and within the Earth. 

\section{Methodology}\label{sec:methodology}

\subsection{Coupled modeling framework}\label{sec:models_proteus}

\noindent To investigate the coupled evolution of Earth's interior and atmosphere during the Hadean, we model the thermal and volatile evolution of a tidally heated magma ocean undergoing progressive solidification. This is achieved using the \texttt{PROTEUS} framework\footnote{\href{https://proteus-framework.org/}{https://proteus-framework.org/}}, which self-consistently couples interior and atmospheric energy transport processes alongside volatile outgassing \citep{Lichtenberg2021,Nicholls_2024_JGRP,Nicholls2025c}.  Tidal heating is imposed through a heuristic parameterization that captures the net power input from tidal dissipation, without relying on spatially resolved stress or viscoelastic models \citep{nicholls_tides_2025}. This allows for exploration of a wide parameter space in tidal power density input while focusing on first-order feedbacks between mantle cooling and atmospheric evolution.

The interior dynamics are modeled with \texttt{SPIDER} \citep{Bower_2018, Bower_2019, Bower2022PSJ}, which is designed to capture the dynamic transition from a fully molten and turbulent magma ocean to solid-state convection dominated by viscous creep, by accounting for energy transport within the mantle through convection, conduction, mixing of phases, and gravitational settling through mixing-length theory. The mantle equation of state is taken to be pure MgSiO$_3$ \citep{WOLF201859}, while the core is assumed to be pure iron with a fixed bulk density, following \citet{Bower_2018}. The mantle is initialized in a fully molten state with an adiabatic temperature profile, consistent with previous studies of magma ocean evolution \citep[e.g.][]{1986EM&P...34..223M, elkins2008linked, hamano2015lifetime, ZAHNLE201574, KrissansenTotton2024}, and similar to the setup adopted by \citet{nicholls_tides_2025}. Magma ocean dynamics models exhibit bottom-up or middle-out crystallization behaviors depending on the adopted liquidus \citep{Bower_2018}, characteristic grain size of cumulates \citep{ballmer_recon_2017, SOLOMATOV201581}, and mixing length parametrization \citep{Bower_2019}. Here, our dynamic mixing length, fixed grain size, and liquidus profile \cite{ANDRAULT2011251} together yield bottom-up fractional crystallization scenarios \citep{elkins2008linked, Lichtenberg2023PP7}.

The critical melt fraction $\phi_{\text{crit}}$ is here set to 0.3 \citep{costa2009model, Bower_2019, SCOTT2006177, Kervazo2021}, below which the mantle's mechanical properties substantially change from liquid to solid; corresponding to the onset of efficient solid-phase tidal heating \citep{farhat2025tides, Hay_2019}. Tidal heating is included through the internal heating term $H$ [\si{W.kg^{-1}.m^{-3}.}] in Equation 1 of the SPIDER methods paper: \citet{Bower_2018}.

Atmospheric structure and evolution are simulated using the \texttt{AGNI} radiative-convective model \citep{Nicholls2025}. \texttt{AGNI} models energy transport in these atmospheres using a mixing-length  parameterization of convection \citep{Robinson_2014} and correlated-k radiative transfer, including Rayleigh scattering \citep{https://doi.org/10.1029/90JD01945, gmd-16-5601-2023}. The numerical method in AGNI determines the atmospheric temperature structure (and energy fluxes) in an energy-conserving manner, allowing for the formation of deep radiative layers, and yielding a realistic solution for planetary thermal evolution \citep{Nicholls_2025_MNRAS}. In this study, clouds and condensation are disabled and no time-dependent photochemistry is modeled; atmospheric composition is isochemical, and determined by thermochemical-solubility equilibrium speciation due to surface in/outgassing at each simulation timestep \citep{Nicholls_2024_JGRP, Nicholls_2025_MNRAS}.

In order to trace the coupled feedback between mantle solidification and atmospheric blanketing in the presence of tidal heating, the rheological front is tracked over time. The rheological front denotes the depth in the mantle at which material transitions from predominantly liquid-like to predominantly solid-like behavior, corresponding to a critical melt fraction at which the mantle’s mechanical and flow properties change regime. Atmospheric blanketing is related to the atmospheric composition, quantified here by the volume mixing ratio (VMR) of each volatile species, and is compared between cases through the net atmospheric energy flux.

\subsection{Tidal Heating}\label{sec:dummy-tides}

\noindent We implement a parametrized tidal heating model where heat is dissipated only in regions of the mantle where its melt fraction $\phi$ is below the critical melt fraction. The heating is scaled linearly with the solid fraction $1 - \phi / \phi_{\text{crit}}$, reaching a maximum in the fully solid case and shutting off when the threshold is exceeded. Tidal heating is applied selectively in the mantle depending on the local melt fraction $\phi$. We emphasize that this prescription of tidal heating is intentionally simplified, in order to preserve the interpretability of the modeled interior–atmosphere evolution, while also remaining agnostic to the particular uncertainties and sensitivities inherent to parametrized rheological models \citep{Renaud_IncreasedT_2018, nicholls_tides_2025, ZAHNLE201574}. Our primary goal is to assess how the presence, magnitude, and radial localization of tidal heating influence magma ocean solidification and atmospheric outgassing, rather than to model tidal dissipation itself in detail. The heating rate per unit mass is given by
\begin{equation}\label{eq:dummy}
    H(\phi) =
    \begin{cases}
    H_0 \left(1 - \dfrac{\phi}{\phi_{\text{crit}}}\right), & \text{if } \phi < \phi_{\text{crit}} \\
    0, & \text{otherwise}
    \end{cases},
\end{equation}
where $\phi$ is the local melt fraction in the mantle; $\phi_{\text{crit}}$ is the critical melt fraction above which tidal heating is disabled; and $H_0 = f(E) \times H_{\text{tide}}$. Here $H_{\text{tide}}$ represents the fixed input parameter defining the maximum possible power density [\si{W.kg^{-1}}], which is varied between simulations to probe the interior--atmosphere response to different heating strengths. The time evolution of tidal heating is modulated by a multiplicative scaling function
\begin{equation}\label{eq:sd_quad}
f(E) = 
\begin{cases}
1, & E \leq E_{\text{crit}} \\
\dfrac{E_{\text{max}}^2 - E^2}{E_{\text{max}}^2 - E_{\text{crit}}^2}, & E_{\text{crit}} < E < E_{\text{max}} \\
0, & E \geq E_{\text{max}}
\end{cases}
\end{equation}
that reflects switching off tidal heating exponentially \citep{Zahnle2007, Heller2021}. This parameterises that, in reality, tidal shutdown could be driven by atmospheric escape or secular evolution of the tidal forcing potential. In particular, planets may experience a decline in tidal heating as satellites migrate outward through tidal evolution, reducing the strength of the tidal forcing \citep{quarles2020application, ZAHNLE201574}, or through the loss or dynamical destabilisation of the satellite due to interactions in multi-planet systems \citep{Payne_2013, Domingos2006}. In addition, abrupt transitions in the tidal response of the system, such as a 'blue sky catastrophe' bifurcation, may also rapidly terminate strong tidal heating \citep{nicholls_tides_2025, PhysRevLett.92.234501, Zhou2013}. In our framework, the exponential decay continues until the integrated dissipated tidal energy reaches $E_\text{max}$, which represents the total energy budget available from the conversion of the Earth's primordial rotational energy through tides. After this energy has been dissipated, strong tidal heating effectively ceases \citep{ZAHNLE201574}. Although tidal dissipation on Earth due to the Moon persists to this day, predominantly through ocean tides, the associated heating rates are orders of magnitude smaller than those considered here \citep[e.g.][]{taguchi2014}. We therefore restrict our analysis to tidal dissipation in the terrestrial mantle during the early stages of the Earth–Moon system. Turning off tidal heating after $E_\text{max}$ is reached enables the determination of the mantle solidification timescale, allowing for quantitative comparison with observationally constrained expectations regarding the Earth's solidification duration. The scaling function $f(E)$ is evaluated for the total dissipated energy $E$ [\si{J}] by tides up-till the current time step during a simulation, relative to a critical value $E_{\text{crit}}$ and a maximum value $E_{\text{max}}$. The critical energy is a dynamic quantity, calculated as 
\begin{equation}\label{eq:e_crit_dyn}
    E_\text{crit} = E_\text{max} - 1.1 \cdot \max{\left(\Delta E\right)},
\end{equation}
where $E_\text{max}$ is the set energy budget, which we vary as an input variable to the code, and $\max{\left(\Delta E\right)}$ represents the largest increase in tidally dissipated energy throughout the simulation. Equation \ref{eq:e_crit_dyn} places the critical energy $1.1$ times the largest increase in energy before the energy budget, making it a dynamic quantity. For large power densities ($10^{-7}$ \si{W.kg^{-1}}), the increase in energy per iteration is much greater than for small power densities ($10^{-10}$ \si{W.kg^{-1}}). Thus, specifying an absolute critical energy value can be problematic. One case can overshoot the budget, while another takes a while to deplete its budget. We include the factor of $1.1$ to prevent overshooting the target value, by ensuring at least one iteration has $E_\text{crit} < E_\text{tot} < E_\text{max}$. As long as the fractional value of the buffer region is small, $(E_\text{max} - E_\text{crit}) / E_\text{max} \ll 1$, the additional numerical scaling factor here has a negligible impact on our model outcomes. For our purposes $E_\text{crit} \lesssim E_\text{max}$. 

The dissipated energy over each iteration $\Delta E$ is calculated, and subsequently summed in order to integrate the dissipated tidal power over time. The energy budget $E_\text{max}$ for tidal heating is fixed at $4 \times 10^{30}$\si{J}, which corresponds to an initial terrestrial angular momentum of $L_\mathTerra^{\text{(ini)}} = 4.1 \ L_\mathTerra$, consistent with an early Earth day length of approximately 6 hours, as supported by previous studies \citep{SPALDING201928, ZAHNLE201574, Canup2020}. The total angular momentum of the Earth–Moon system is assumed to be equal to its present-day value ($L_{EM}^\text{(ini)} = L_{EM}$), in line with canonical formation models \citep{Canup2020}. This places the Moon with its present mass and density at the Roche limit (2.9 $R_\mathTerra$; \citealt{ZAHNLE201574}) shortly after formation. 

Given that the contributions of radioactive heating and stellar evolution to the planet's energy budget are negligible in the scenarios modeled here, the tidal dissipation durations scale linearly with the target energy budget $E_\text{max}$. Therefore, here we limit our analysis to one value of $E_\text{max}$. 

The above described formalism models a scenario in which tidal heating is initially small when the mantle is molten, increasing as solidification proceeds, then shutting off again when $E_\text{max}$ is exceeded. The simulations terminate when either the mantle reaches global melt fraction $\leq 0.005$ while the surface tidal heating flux $\leq 0.1$ \si{W.m^{-2}}, or when the simulated time exceeds 1 billion years.

\subsection{Planetary Parameters}
\label{sec:cases}

\noindent We take the planet's semi-major axis as 1.0 AU with an orbital eccentricity of 0.0167. Stellar evolution is modeled self-consistently using \texttt{MORS} \citep{johnstone2021}, based on luminosity and radius tracks from \citet{Spada_2013}. A current stellar age of 4.567 Gyr is adopted for generating the input spectra. To approximate global radiative forcing with a single 1D atmospheric column, we assume a fixed solar zenith angle of $48.19^\circ$ and apply an instellation scale factor of 0.375 to account for day-night averaging \citep{hulstrom1985spectral, cronin_zenith_2014}.

To assess the possible range of interaction between tidal heating and the early Earth's atmosphere, a wide parameter space is explored. The tidal power density is to be varied; different Moon-forming and Earth-Moon-evolution scenarios predict surface power fluxes ranging from 100 to 0.1 W m$^{-2}$ \citep{Zahnle2007,Canup2020,Heller2021}. These values are roughly equivalent to power densities between $10^{-7}$ and $10^{-10}$\,\si{W.kg^{-1}}. Hence, we consider tidal power density values of $10^{-7}, 10^{-8}, 10^{-9}, 10^{-10}$ and 0\,\si{W.kg^{-1}}, where the latter represents a baseline scenario without tidal heating. These power densities place a lower bound on the mantle solidification timescale (with $L_{EM}^\text{(ini)} = L_{EM}$) under the assumption that it is fixed with time. Tidal power density is expected to decrease with time as the Earth spins down and the lunar orbit recesses \citep{ZAHNLE201574, Canup2020, Heller2021, SPALDING201928}. Given this, smaller tidal power densities in our formalism likely occured at a later stage in the evolution, and represent an upper bound to the mantle solidification timescale compared to the (more realistic) scenario of varying tidal power density. Similarly, larger tidal power densities are more plausible shortly after the moon-forming impact, and represent a lower bound to the mantle solidification timescale \citep{ZAHNLE201574, KORENAGA2025116743}. In parameterizing tidal heating as a constant quantity, we are not sensitive to the particular solution to the tidal equations or a model of mantle rheology \citep{driscoll2015tidal, Henning_2009}.

The oxidation state of the Hadean Earth is poorly constrained, with early reducing conditions often assumed but geochemical evidence indicating a relatively oxidized mantle by $\sim 4.35$ Ga \citep{FrostMcCammon2008,Schaefer2017,Trail2011}. Given this, the oxygen fugacity ($f$O$_2$) is varied from $-4$ to $+4$ log units relative to the iron-w\"ustite buffer ($\Delta \text{IW}$). This offset relative to IW is held constant throughout each simulation, which presents another bound on the degassing properties of the mantle and feedback with the atmosphere. In reality, we would expect the $f$O$_2$ of the upper mantle to increase in time due several processes, including core formation \citep{Deguen2014}, iron disproportionation \citep{Schaefer2024}, and H-escape to space \citep{catling2020archean}. The impact of element abundances on the atmospheric composition is investigated by varying the initial inventory of volatile elements in the mantle+atmosphere system. We do not model or account for partitioning into the metal core and hence our stated abundances are below the expected bulk planet abundance. We vary the hydrogen abundance from $2$ to $5$ modern day Earth's ocean mass equivalents (1 Earth ocean of H$_2$O $\sim 10^{21} \si{kg}$). We assume a nominal case composed of 3 modern Earth oceans, a C/H ratio of 1, S/H ratio of 2, and N/H ratio of 0.5 \citep{wang_Theeleme_2018}. We do not account for volatile partitioning in the solid phase of the mantle during crystallization nor atmospheric escape to focus our results on the tide-atmosphere feedback. 

A key metric and finding from our simulation outcomes are lasting periods of `global radiative equilibrium' (GRE). These epochs represent periods of quasi-steady-states within the Earth interior-atmosphere system, where the net amount of energy transported through the atmosphere to space is equal to the total power dissipated within the planet's interior plus the (comparably minor) energy flux received from the young Sun. These GRE epochs represent temporary stable climate states where the net energy flux of the planet is in equilibrium. In our models, the Earth exits GRE epochs only when tidal heating shuts-down according to the total energy budget described above. In reality, tidal potentials, and therefore $H_{\text{tide}}$, are not constant in time but decay as the Earth–Moon system evolves through coupled tidal–orbital dynamics. We speculate on the feasibility of long-lasting GRE epochs in such coupled systems in Section~\ref{sec:GREdisc}.

\subsection{Lunar Orbital Recession}

\noindent Lunar orbital recession reduces tidal power density by weakening the Earth–Moon gravitational interaction. While the Moon’s orbital evolution is not explicitly modeled here, the phase-lag model allows a relation between tidal power density and the lunar semi-major axis to be established \citep{greenberg2009frequency, efroimsky2013tidal}.

The torque exerted by the Moon on the Earth is given by
\begin{equation}\label{eq:torque_def}
N = \frac{9}{4} G \frac{A_\mathTerra^5}{Q'_\mathTerra}\frac{m_\text{M}^2}{R^6},
\end{equation}
where $G$ is the gravitational constant, $A_\mathTerra$ Earth’s radius, $Q'_\mathTerra = (2.5/h) Q_\mathTerra$ the modified tidal quality factor, $h$ the Love number, $m_\text{M}$ the Moon’s mass, and $R$ the Moon’s orbital semi-major axis \citep{Goldreich1966}. The tidal $Q$ parameter is defined as
\begin{equation}\label{eq:tidal_Q}
Q^{-1} = \frac{1}{2 \pi E_0} \oint - \frac{dE}{dt} dt,
\end{equation}
where $E_0$ is the total potential energy available for dissipation during one lunar revolution. During a constant tidal heating phase (GRE), this becomes
\begin{equation}\label{eq:tidal_Q_GRE}
Q^{-1}_{\text{GRE}} = \frac{1}{2 \pi}\frac{H_\text{tide}^\text{eff}}{H_\text{tide}},
\end{equation}
with effective power density $H_\text{tide}^\text{eff}$ defined as
\begin{equation}\label{eq:power_mantle}
H_\text{tide}^\text{eff} = P/M_{\mathTerra, \text{mantle}},
\end{equation}
where $P = dE/dt$ [\si{W}] is the tidal power and $M_{\mathTerra, \text{mantle}}$ is the mantle mass.

The torque is also related to Earth’s rotational angular momentum loss
\begin{equation}\label{eq:torque_ang}
N = -\frac{dL_\mathTerra}{dt} = -\frac{d}{dt}\left[\frac{dE}{d\omega}\right].
\end{equation}
Assuming uniform angular deceleration, and constant $dE/dt$ during GRE, this yields
\begin{equation}\label{eq:torque_ang_simple}
N = \left(\frac{dE}{dt}\right)\left[\frac{d^2 \omega}{d t^2}\right] \left(\frac{d \omega}{d t}\right)^{-2}\sim \left(\frac{dE}{dt}\right) \frac{1}{\Delta \omega},
\end{equation}
where $\Delta \omega$ is the change in Earth’s spin rate over time.

Equating expressions for $N$ and simplifying via equations \ref{eq:tidal_Q_GRE} and \ref{eq:power_mantle}, the lunar orbital radius during GRE becomes
\begin{align}\label{eq:R_GRE_Q}
R_\text{GRE} &\approx \left[ \Delta \omega \frac{9G}{20 \pi} \frac{m_\text{M}^2 A_\mathTerra^5}{M_{\mathTerra, \text{mantle}}} \frac{h}{H_\text{tide}} \right]^{1/6}.
\end{align}
This scaling illustrates the dependence of the lunar semi-major axis on tidal power density. However, it assumes static equilibrium tides and does not capture frequency-dependent or viscoelastic effects \citep{farhat2025tides, Efroimsky_2009}, and thus serves as a simplified, order-of-magnitude estimate.

\section{Results}\label{sec:results}

\begin{figure}[t!]
  \centering
  \includegraphics{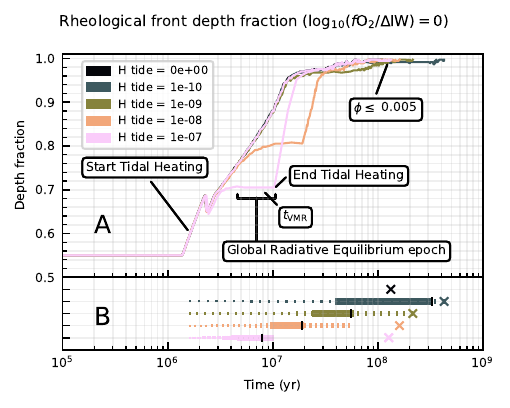}
  \vspace{-0.6cm}
  \caption{Simulation results for nominal abundance and $f$O$_2 = \Delta \mathrm{IW} + 0$ across all tidal power densities (see Section \ref{sec:cases}). The horizontal axis shows time [\si{yr}] since the Moon-forming impact on a logarithmic scale. Panel A: evolution of the rheological front (vertical axis: mantle depth fraction). Panel B: evolution of tidal dissipation for each case (vertical axis: case label). Dotted lines trace cumulative tidal energy dissipation; marker size scales with total dissipated energy. Solid lines indicate tidal heat-supported GRE; black markers ($\pmb |$) denote the time $t_\mathrm{VMR}$ at which we extract the volume mixing ratio for plotting in Figure~\ref{fig:VMR_fo2+0_evo}. Note that the particular choice of $t_\mathrm{VMR}$ is arbitrary so long as it lies within the temporal range of GRE, indicated by the solid lines in the bottom panel. Crosses ($\times$) mark mantle solidification ($\phi < 0.005$).}
  \label{fig:Rheo_tide_timeline}
\end{figure}

\noindent A central finding from our simulations is that tidal power densities between $10^{-10}$ and $10^{-7}$ W kg$^{-1}$ are able to support stable global radiative equilibrium (GRE) epochs for extended periods. The early release of H$_2$ in reducing cases allows strong tidal heating to maintain steady-states at the largest observed melt fractions in our models. The late release of H$_2$O in oxidizing cases produces a relatively stronger greenhouse, allowing weaker tides to maintain magma oceans. The atmospheric composition in reducing cases does not change significantly, as such weaker tides do not benefit from additional blanketing in these cases. Overall, this shows that due to the atmospheric greenhouse properties, tidal heating is most limited in reducing cases for strong tides, whilst heating due to weaker tides is more limited in oxidizing cases. We will now go into the details, starting with the rheological front, then atmospheric evolution, and atmospheric composition. Finally, we provide a sensitivity study of our findings to the initial hydrogen abundance.

\subsection{Rheological front}\label{sec:timeline+rheo}

\begin{figure*}[tbh!]
  \centering
  \includegraphics{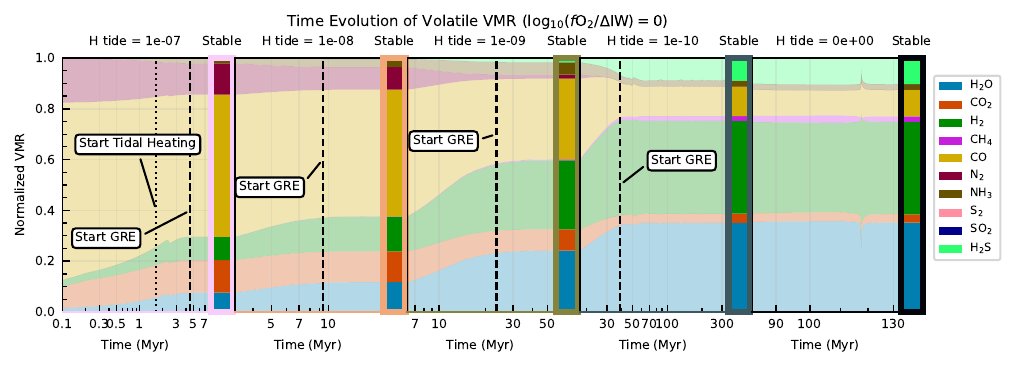}
  \vspace{-0.5cm}
  \caption{Simulated atmospheric compositions (expressed as normalized volume mixing ratio, VMR) for cases with nominal elemental abundances and $f$O$_2 = \Delta \mathrm{IW} + 0$,  across all considered tidal power densities (see Section \ref{sec:cases}). The horizontal axis denotes the time after model initialization. The horizontal axis represents five axis stitched together, each leading up to their corresponding $t_\mathrm{VMR}$ (Figure \ref{fig:Rheo_tide_timeline}B), and for the case $H_\text{tide}=0$, the time of solidification. Cases transition smoothly across this plot because they all attain similar compositions in their evolution, only deviating from the case without tidal heating when tides maintain specific non-zero melt fractions corresponding to $H_\text{tide}$. Weaker tides become active later, and so while these cases attain similar compositions as their stronger tidally heated counterparts, the time at which they do so varies. The dotted vertical lines in the shaded region represent the start of GRE epochs, demonstrating that the atmosphere composition remains unchanged in the absence of escape processes.}
  \label{fig:VMR_fo2+0_evo}
\end{figure*}

\noindent Figure \ref{fig:Rheo_tide_timeline}A shows the time evolution of the rheological front for cases with $f\text{O}_2 = \Delta\text{IW}+0$ at different power densities. The rheological front starts at the bottom of the mantle at a radius fraction of $0.55$ corresponding to the core-mantle boundary. During the first million years the mantle is in a liquid state. When the lower layers of the mantle start to cool, crystallization begins, and the rheological front moves upwards. The magma becomes increasingly more viscous, and only when the local melt fraction becomes less than 30\% ($\phi_\text{crit}$) does tidal heating begin. The melt fraction varies throughout the mantle, so tidal heating acts locally. The minor discontinuity around depth fraction equal to 0.7 at about 2--3 Myr arises from the solidification of the lowermost layer of the mantle, which influences the rate of transfer from the core to the mantle \citep{Nicholls_2024_JGRP, Bower_2019, Bower_2018}. 

All cases in Figure \ref{fig:Rheo_tide_timeline} that include tidal heating show a slow increase in thermal energy dissipation starting after 1.5 million years of evolution, indicated by the start of the dotted lines in the timeline plot of the lower panel (Figure \ref{fig:Rheo_tide_timeline}B). From this point, the progress of the rheological front is slowed down until reaching a tidally supported GRE steady-state. The atmospheric composition is constant during this period. We find that larger tidal power densities are able to reach their radiative equilibrium state sooner and maintain larger melt fractions (Figure \ref{fig:Rheo_tide_timeline}, panel A); these stronger tides dissipate energy more rapidly, yielding shorter GRE durations given our fixed energy budget. Overall, it is evident that the imposed internal heating rates are able to significantly alter the thermal evolution of the modeled planet.

\subsection{Atmospheric evolution}\label{sec:VMR_GRE}

\begin{figure}[tbh]
  \centering
  \includegraphics{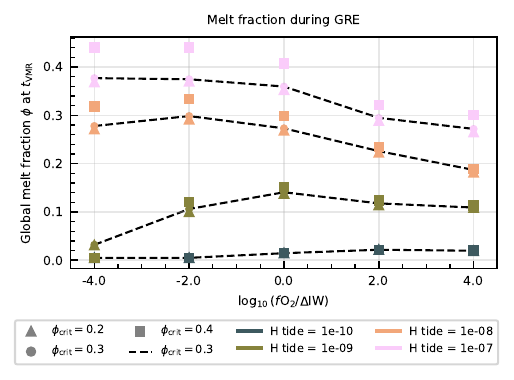}
  \vspace{-0.6cm}
  \caption{Global melt fraction $\phi$ during global radiative equilibrium (at $t_\mathrm{VMR}$) as a function of $\log_{10}(f\mathrm{O}2/\Delta$IW) for a range of tidal power densities $H_\text{tide}$ (marker color) [\si{W.kg^{-1}}] and critical melt fraction $\phi_\text{crit}$ (marker shape).}
  \label{fig:melt_spread}
\end{figure}

\noindent Solidified layers in the mantle are entirely depleted of volatiles in our model. During bottom-up crystallization, volatiles move to the outer shells of the mantle and are subsequently outgassed at the surface due to the decreasing volume of melt. The atmosphere is composed of volatiles outgassed in equilibrium from the mantle. The relative abundance of volatiles in the atmosphere changes with time, an effect dictated by solubility of volatiles through empirical solubility laws and by equilibrium chemistry in the atmosphere \citep{Nicholls_2024_JGRP, Lichtenberg2021, SOSSI2023117894, Bower2022PSJ}. As the melt fraction decreases, only the most highly soluble volatiles (notably H$_2$O) remain dominantly in the mantle. Ultimately, all volatiles are forced out of the mantle at complete solidification, independent of their solubility. 

In Figure \ref{fig:VMR_fo2+0_evo} the evolution of atmospheric composition are shown as volume mixing ratios (VMRs) for cases with nominal abundance and $f$O$_2 = \Delta \mathrm{IW} + 0$. The atmospheric composition evolves similar to the case without tides, until tidal heating starts ($\sim1.5$\,Myr). The case with $H_\text{tide}=10^{-7}$ \si{W.kg^{-1}} diverges from the no-tides case first, reaching GRE after $4.8$\,Myr, indicated by the dashed vertical line (Figure \ref{fig:VMR_fo2+0_evo}). Slower progression of the rheological front due to tidal heating acts to slow down the atmospheric compositional evolution; $H_\text{tide}=10^{-8}$ \si{W.kg^{-1}} reaches the same state at an earlier time due to weaker tides. The other cases ($H_\text{tide}=10^{-9}$ and $=10^{-10}$ \si{W.kg^{-1}}) reach GRE after 22\,Myr and 40\,Myr, respectively. Figure~\ref{fig:VMR_fo2+0_evo} shows that the atmospheric composition evolves in a similar manner across all cases considered, stabilizing once the solidification front stagnates as a result of tidal dissipation. Weaker values of $H_{\text{tide}}$ allow the solidification front to progress further upward through the mantle before stalling during GRE epochs, corresponding to smaller whole-mantle melt fractions and proportionally more degassing of soluble volatiles (notably $\rm H_2O$). Longer solidification timescales in weakly tidally heated cases result from reduced energy dissipation rates for weaker tides, and cooler, less extended atmospheres that radiate heat less efficiently. In contrast, strong tidal heating sustains hotter atmospheres that enhance cooling and accelerate solidification. 

\begin{figure*}[t!]
  \centering
  \includegraphics{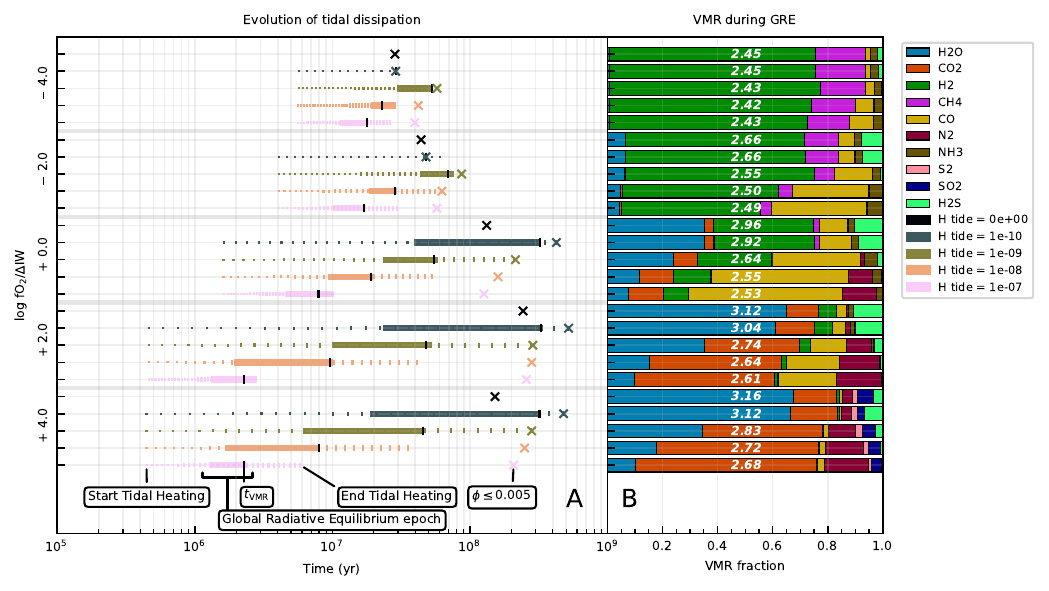}
  \vspace{-0.6cm}
  \caption{Evolution of nominal-composition cases across the full range of $f$O$_2$ and tidal power densities simulated (see Section \ref{sec:cases}). Panel A (left): stages in planet's lifetime after model initialization. Dotted lines trace cumulative tidal heat dissipation; marker size increases with total dissipated energy. Solid lines indicate tidally-supported GRE states; black vertical bars ($\pmb |$) mark $t_\mathrm{VMR}$. Crosses ($\times$) denote solidification, defined at $\phi < 0.005$. Panel B (right): volume mixing ratios of volatiles at $t_\mathrm{VMR}$ and white numbers show corresponding $\log_{10}$ total surface pressure [bar].}
  \label{fig:timelines_VMR}
\end{figure*}

Figure \ref{fig:melt_spread} also shows melt fractions for cases with $\phi_\text{crit} = \{0.2, 0.4\}$. At low $H_\text{tide}$, the critical melt fraction has little effect on $\phi$. Increasing $\phi_\text{crit}$ only affects high $H_\text{tide}$, producing larger melt fractions, especially in reducing mantles. These trends emphasize the central role of $f$O$_2$ and the feedback between atmospheric composition and $H_\text{tide}$ in controlling the thermal evolution of the mantle–atmosphere system: reducing conditions promote early, efficient outgassing and hence decreased heat flux to space, while oxidizing conditions delay atmospheric buildup.

\subsection{Atmospheric composition}
\label{sec:eq_chem}

\noindent To capture the effects that different mantle and hence atmospheric compositions can have on tidal heat dissipation, the oxygen fugacity is varied from $-4$ to $+4$ relative to the iron-w\"ustite buffer. The resulting timelines (Section \ref{sec:timeline+rheo}) and corresponding volume mixing ratios during GRE states for each tidal power density are shown in Figure \ref{fig:timelines_VMR}. 

Figure \ref{fig:timelines_VMR} highlights clear trends which emerge between our most oxidizing ($\Delta \mathrm{IW} + 4$) and reducing ($\Delta \mathrm{IW} - 4$) cases. Reducing cases yield H$_2$- and CH$_4$-dominated atmospheres throughout. In contrast, oxidizing cases are dominated by H$_2$O, CO, CO$_2$, and N$_2$, consistent with redox-controlled volatile speciation \citep{Lichtenberg2021, Nicholls_2024_JGRP, sossi2020redox, Bower2022PSJ}.

Given the dependence of atmospheric composition on mantle solidification (Section \ref{sec:VMR_GRE}), the variation in global melt fraction for a given tidal power density (Figure \ref{fig:melt_spread}) explains the differences in volume mixing ratios (Figure \ref{fig:timelines_VMR}, panel B). As is also shown in Figure \ref{fig:VMR_fo2+0_evo}, less soluble volatiles (e.g., CO, CO$_2$, N$_2$) are released at high melt fractions, while more soluble species (primarily H$_2$O) degas near complete mantle solidification. CH$_4$ is only present at low $f$O$_2$. Sulfur speciation is controlled both by $f$O$_2$ and $H_\text{tide}$ through the temperature dependence of H$_2$S formation \citep{Nicholls2025c}. At $f$O$_2 = \Delta$IW$+4$ and high temperatures, SO$_2$ dominates, but at lower $H_\text{tide}$ (e.g., $10^{-10}$ \si{W.kg^{-1}}), H$_2$S becomes more abundant. At solidification, SO$_2$ dominates in oxidizing conditions due to greater oxygen availability, while H$_2$S dominates at $f$O$_2 \leq \Delta$IW$+2$. Nitrogen speciation is controlled by the sensitivity of N solubility on $f$O$_2$ \citep{LIBOUREL20034123,Shorttle2024} and the temperature dependence of the formation of NH$_3$ from H and N \citep{Nicholls2025c}: N$_2$ is the dominant carrier at oxidizing conditions, while it is NH$_3$ at reducing conditions. Figure \ref{fig:timelines_VMR} (panel B) suggests that intense early tidal heating depleted atmospheric H$_2$O while the mantle remained molten, yielding atmospheres rich in N$_2$ and CO$_2$. In contrast, lower tidal heating would have increased atmospheric NH$_3$ and H$_2$S, adding a disequilibrium effect to atmospheric composition.
\begin{figure}[tbh]
    \centering
    \includegraphics{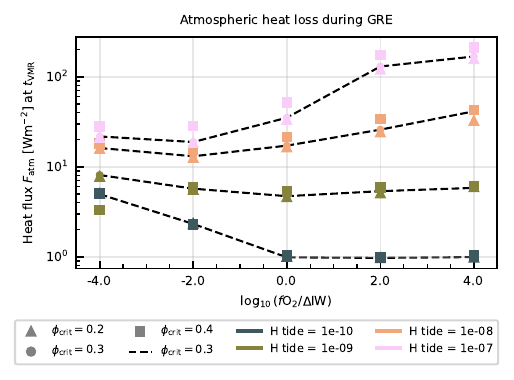}
    \vspace{-0.6cm}
    \caption{Net atmospheric energy flux, by which the planet loses heat to space balanced against tides during GRE. Plotted at time $t_\mathrm{VMR}$ as a function of $\log_{10}(f\mathrm{O}2/\Delta$IW) for different tidal power densities $H_\text{tide}$ and $\phi_\text{crit}$.}
    \label{fig:atm_heat_flux_abs}
\end{figure}
In Figure \ref{fig:timelines_VMR}B, the volume mixing ratios in the most reduced scenarios ($f\text{O}_2 = \Delta\text{IW}-4$) appear insensitive to tidal heating. Figure \ref{fig:atm_heat_flux_abs} confirms that atmospheric heat loss at GRE is also largely insensitive to tidal power in these cases, reinforcing the compositional similarity across tidal scenarios.
\begin{figure*}[tbh!]
    \centering
    \includegraphics{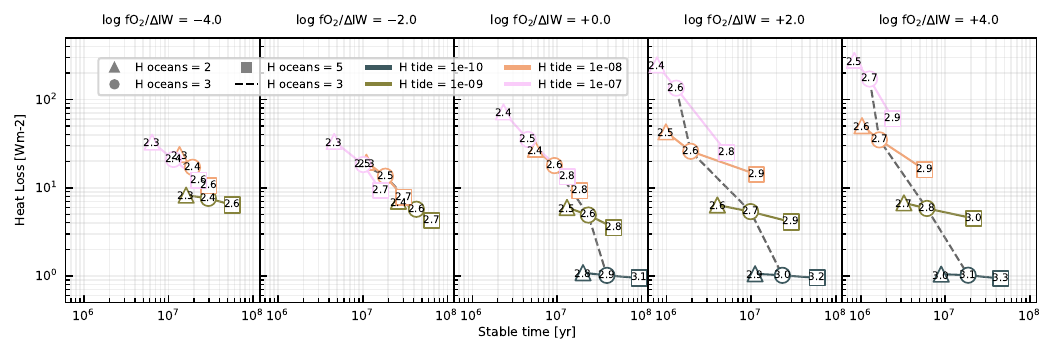}
    \vspace{-0.6cm}
    \caption{Net atmospheric energy flux to space during GRE versus onset time of the GRE epoch [\si{yr}], numbers show corresponding $\log_{10}$ total surface pressure [bar]. Increasing hydrogen abundance reduces radiative cooling to space, extending low melt-fraction magma ocean lifetimes and postponing GRE epochs. Tidal dissipation within the mantle is the dominant source of atmospheric heat fluxes in the majority of these scenarios. At large $H_\text{tide}$, one can consider the tidal and total heat fluxes to be equivalent; for the weakest-heating case of $H_\text{tide} = 10^{-10}$, the tidal dissipation contributes 80\% of the total heat flux. The remaining 20\% comes from radiogenic heating.
}
    \label{fig:sens_H_atm}
\end{figure*}
Atmospheres formed above reducing interiors initially cool more slowly, but consistently solidify earlier than their more oxidized counterparts (Figure \ref{fig:timelines_VMR}A). This is because strongly reducing atmospheres permit high tidal power densities ($\geq 10^{-8}$ \si{W.kg^{-1}}) to sustain high melt fractions (Figure \ref{fig:melt_spread}), where tidal dissipation is inefficient. Weaker tidal inputs ($\leq 10^{-9}$ \si{W.kg^{-1}}) cannot maintain large melt fractions. Since the atmospheric composition remains nearly constant across reducing cases, radiative heat loss also stays roughly constant (Figure \ref{fig:atm_heat_flux_abs}). The CH$_4$/CO ratio does evolve from CO- to CH$_4$-dominated atmospheres as the surface cools, but this shift is minor and less pronounced with higher hydrogen inventories, and can thus be neglected.

Together, these effects in reducing cases imply that only tidal power densities $\geq 10^{-8}$ \si{W.kg^{-1}} can offset atmospheric losses during GRE for reducing conditions at $f \text{O}_2 \leq \Delta\text{IW} -2$. Weaker tides are insufficient to stabilize thermal evolution at low melt fractions, and no greenhouse feedback compensates for this. In contrast, oxidizing atmospheres ($f\text{O}_2 > \Delta\text{IW}+0$) respond more dynamically to tidal input (Figure \ref{fig:timelines_VMR}B): late-stage H$_2$O release enhances greenhouse warming as the melt fraction declines, reducing the atmospheric energy flux more strongly with increasing $f\text{O}_2$ (Figure \ref{fig:atm_heat_flux_abs}). This allows weaker tides to sustain higher melt fractions (Figure \ref{fig:melt_spread}) and delays solidification (Figure \ref{fig:timelines_VMR}A).

This behavior persists across different critical melt fractions. Figure \ref{fig:atm_heat_flux_abs} shows that increasing $\phi_\text{crit}$ increases atmospheric heat loss in oxidized cases, especially at high $H_{\text{tide}}$. However, when $H_{\text{tide}} \leq 10^{-9}$\,\si{W.kg^{-1}} and the rheological front is near the surface (Figure \ref{fig:Rheo_tide_timeline}A), the additional heating in the outermost layers of the mantle becomes negligible (larger melt fraction). At low melt fractions, the influence of $\phi_\text{crit}$ weakens, and any differences between choices of $\phi_\text{crit}$ are further nullified by an enhanced greenhouse effect from late H$_2$O release (Figure \ref{fig:timelines_VMR}B).

\subsection{Hydrogen Abundance}

\noindent Hydrogen abundance strongly influences mantle solidification timescales. We found that increasing hydrogen abundance from 2 to 5 Earth oceans (for zero tidal heating) delays magma ocean solidification by tens of Myr. Higher hydrogen content decreased atmospheric heat loss (Figure \ref{fig:sens_H_atm}) by increasing absolute gas abundances and surface pressure, thus enhancing greenhouse warming and maintaining higher surface temperatures and melt fractions. This is in line with previous numerical simulations \citep{Nicholls_2024_JGRP, hamano2015lifetime} and expectations from analytical atmosphere models \citep{pierrehumbert2010principles, guillot2010radiative}.

\begin{figure}[t!]
    \centering
    \includegraphics{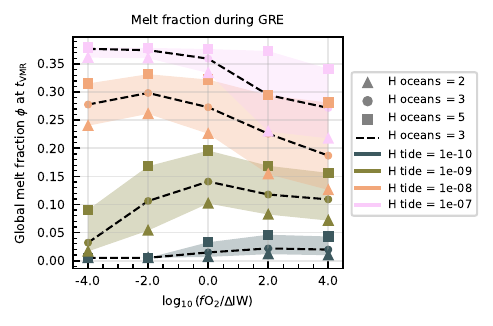}
    \vspace{-0.6cm}
    \caption{Global melt fraction during the GRE epoch. Higher hydrogen abundances correspond to more extensive melting. Similar to Figure \ref{fig:melt_spread}, late-stage H$_2$O release enhances greenhouse warming as melt fraction declines, reducing atmospheric energy flux more strongly at higher $f$O$_2$ and allowing weaker tides to maintain higher melt fractions.}
    \label{fig:melt_H}
\end{figure}

Under reducing conditions ($f$O$_2 = \Delta$IW$-4$), added hydrogen promotes CH\textsubscript{4} formation over CO, particularly at lower surface temperatures. At high $H_\text{tide}$ ($10^{-7}$ \si{W.kg^{-1}}), global melt fractions are insensitive to hydrogen at $f$O$_2 < \Delta$IW$+0$, since the melt fractions already approach the critical melt fraction (Figure \ref{fig:melt_H}), limiting tidal heat feedback. For intermediate $H_\text{tide}$, global melt fractions during GREs are most hydrogen-sensitive. The overall increase in atmospheric H\textsubscript{2}O in oxidizing cases reduces tidal surface fluxes ($200 \rightarrow 50$ [\si{W.m^{-2}}], Figure \ref{fig:sens_H_atm}) and yields the greatest sensitivity of the global melt fraction to the total hydrogen content (Figure \ref{fig:melt_H}). 

\section{Discussion}\label{sec:discussion}

\subsection{Mantle solidification timescales}
\noindent Initial simulations in the base case of no tidal heating, with $f$O$_2$ ranging from $\Delta$IW$-4$ to $\Delta$IW$+4$, show that an Earth-like planet undergoes complete mantle solidification within 150\,Myr after a potential Moon-forming impact. These timescales align broadly with previous results in the literature (e.g. \citet{1985LPSC...15..545A, 1986EM&P...34..223M, hamano2015lifetime, ZAHNLE198862, Hamano2013, refId1, 2021AsBio..21.1325B, KrissansenTotton2024}). However, some works find a shorter solidification timescale compared to our results (e.g. \citet{https://doi.org/10.1002/jgre.20068, sahu2025unveilinginteriorstructurethermal}), principally because of different atmospheric models and an often lower inventory of volatiles considered. We identified a dependence of mantle solidification timescales on the oxygen fugacity of the mantle \citep{Nicholls_2024_JGRP}. More reducing cases solidify faster than the oxidized cases; generally taking only 30\,Myr, while cases with $f$O$_2 \geq \Delta \mathrm{IW} +0$ may take up to 150 Myr to solidify. These differences are consistent with literature values for hydrogen-rich atmospheres (more reducing), as well as water and carbon-oxide rich atmospheres (more oxidizing) \citep{Lichtenberg2021, 1985LPSC...15..545A, 1986EM&P...34..223M, hamano2015lifetime, ZAHNLE198862, Hamano2013, 2021AsBio..21.1325B, KrissansenTotton2024, refId1}.

Let us now consider cases with parametrized tidal heating incorporated into the modeled evolution, again under different atmospheric scenarios. Heating from tides generally extends the solidification time of the Earth's mantle (Figure \ref{fig:timelines_VMR}A). Consistent with Section~\ref{sec:cases}, we find that the weakest tidal forcing scenarios lead to the largest increase in solidification time, while the highest tidal power density scenarios are closest to the scenario without tidal heating. This decrease in solidification time with increasing tidal power density for the tidally-active scenarios is a result of slower energy dissipation due to smaller tidal heating rates imposed by smaller tidal power densities: strongly melted and hot cases lead to increased atmospheric scale heights that cool faster more efficiently. At lower heating rates, energy is deposited more evenly in the mantle without a large increase in energy flux through the atmosphere. The differences in solidification time between tidal power densities are the largest under more oxidizing regimes. In reducing regimes, cases with $H_\text{tide}=10^{-10}$ \si{W.kg^{-1}} solidify around the same time as the cases without tides (Figure \ref{fig:timelines_VMR}A). 

The tidal dissipation timescales found for cases with $H_\text{tide} \geq 10^{-9}$ \si{W.kg^{-1}} (Figure \ref{fig:timelines_VMR}A) fall within or below the tidal dissipation timescale of $\sim$100 Myr proposed by \citet{Zahnle2007} and \citet{tidestimes}. However, tidal dissipation/solidification timescales up to 330 Myr in our models for $10^{-10}$ \si{W.kg^{-1}} far exceed 100 Myr. Following Section~\ref{sec:cases}, these results support the interpretation that higher tidal power densities yield lower bounds on realistic mantle solidification timescales, while lower values offer upper bounds. The overall alignment with the literature suggests that our adopted parameter space bounds a range of plausible scenarios. To extend on these estimates, future work should allow $H_\text{tide}$ to evolve self-consistently over time based on the rheological properties of the mantle, to provide more accurate estimates of the evolution of the early Earth--Moon system.

\subsection{Atmospheric composition}\label{sec:atm_compos_dis}

\noindent Differences in mantle solidification time that arise between the cases at different tidal power densities and oxygen fugacities follow from variations in atmospheric composition. Based on the volume mixing ratios plotted in Figure \ref{fig:timelines_VMR}B, we find that highly oxidized mantles ($f\text{O}_2 =\Delta$IW$ +4$) mainly outgas oxidized species, such as CO$_2$ and H$_2$O. Moving to $f\text{O}_2 = \Delta$IW$ +0$, the balance shifts away from CO$_2$ and towards CO; whilst H$_2$O is partially converted to H$_2$. There is unanimous absence of atmospheric H$_2$O under the most reducing cases that we have considered ($f\text{O}_2 =\Delta$IW$ -4$), where instead H$_2$ is the dominant species alongside CH$_4$. These findings are consistent with previous experimental and theoretical results in the literature \citep{Schaefer2017, GAILLARD2022117255, salvador2023magma, sossi2020redox, Bower2022PSJ, seidler_Impactof_2024, Boer2025}.  

Comparing the atmospheric composition at different tidal power densities with each other gives insight into the temporal evolution of the atmospheric composition (Figure \ref{fig:timelines_VMR}). We find that oxidizing atmospheres are initially dominated by less-soluble species (CO$_2$, N$_2$; \citet{Lichtenberg2021}). When the mantle approaches solidification H$_2$O is preferentially outgassed. At oxygen fugacities of $f$O$_2 = \Delta$IW$+ 0$, CO is the initial dominant species paired with small contributions of H$_2$O, CO$_2$, H$_2$, and N$_2$. As a new finding from the coupled feedback of tides and outgassing, we find that relatively weak tides lead to enhanced atmospheric abundance of H$_2$S and NH$_3$ due to the temperature dependence of their equilibrium reaction rate. This suggests a potential path to probe tidal disequilibrium processes on Hadean Earth analog exoplanets with space telescopes \citep{Bonati19,Cesario24}.

The atmospheric compositions plotted in Figure \ref{fig:timelines_VMR}B can be compared to the corresponding atmospheric energy fluxes shown in Figure \ref{fig:atm_heat_flux_abs} to quantify the greenhouse effect of the atmosphere, given the same instellation flux. We find that the oxidizing atmospheres evolve to become extremely opaque to radiation from the surface. The reducing atmospheres are initially more effective at blanketing the surface than oxidizing cases, due to the rapid formation of H$_2$ dominated atmospheres which have large infrared opacity from the H$_2$ self-collisional continuum \citep{pierrehumbert2010principles}. \citet{Lichtenberg2021} previously found that large H$_2$ abundances make for a stronger greenhouse gas than CO$_2$ at high pressures and fixed atmospheric composition. However, once the H$_2$O degasses from the mantle in the oxidizing cases, these atmospheres are stronger heat flux blankets.

We find that oxidizing cases dissipate energy at rates up to 200 [\si{W.m^{-2}}] at global melt fractions as large as 0.28. This state is maintained by tidal power densities of $10^{-7}$ \si{W.kg^{-1}} \citep{2021AsBio..21.1325B, nicholls_tides_2025}. Reducing cases display smaller net atmospheric energy fluxes ranging up to 30 \si{W.m^{-2}} at global melt fractions as large as 0.38 maintained by tidal power densities $= 10^{-7}$ \si{W.kg^{-1}} \citep{refId1}. At smaller melt fractions tidal heating becomes more efficient via Equation \ref{eq:dummy}, however the surface energy fluxes decrease to $<1$ \si{W.m^{-2}}, ($H_\text{tide}$ $< 10^{-10}$ \si{W.kg^{-1}}). These findings are mostly in agreement with literature values of tidal heat flux ranging from 0.1 to 100 \si{W.m^{-2}}  \citep{Zahnle2007, Canup2020, Heller2021}. Given that atmospheric composition in reducing cases is mostly agnostic to the effects of tidal heating, it follows that our values for these cases agree with literature results which have assumed a fixed atmosphere. However, we find substantially larger tidal heating rates for oxidized cases (i.e. $\gtrsim 100$ \si{W.m^{-2}}). Our results show a balance between total tidal power density and mantle melt fraction at a steady state because larger tidal heating rates support larger global melt fractions (Figure \ref{fig:melt_spread}). 

Our results display small sensitivity to the critical melt fraction parameter. Mainly reducing cases show an increase in melt fraction due to an opaque atmospheric rich in H$_2$ which is relatively insensitive to the melt fraction through its poor solubility. In oxidizing cases the greenhouse effects are weak at large melt fractions because of the dissolution of H$_2$O which allows efficient cooling to space. These results are consistent with results from \citet{KORENAGA2023115564}, who showed that under a dense CO$_2$-rich atmosphere the critical melt fraction does not significantly alter lunar recession rates. 

\subsection{Tidal Efficiency}
\begin{figure}[t!]
    \centering
    \includegraphics{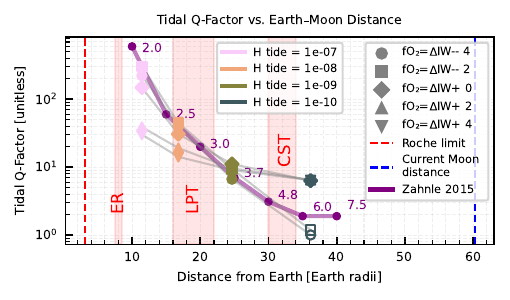}
    \vspace{-0.6cm}
    \caption{The distance to the Moon and the tidal dissipation parameter $Q$ of the Earth after a canonical Moon-forming impact. Using Equation \ref{eq:tidal_Q_GRE}, the tidal $Q$ parameters (efficiency of tides) corresponding to the individual simulations are calculated, subsequently Equation \ref{eq:R_GRE_Q} is used to estimate the lunar orbital recession at the tidal power densities corresponding to the $Q$-parameters. Data points for $f$O$_2 = \Delta \mathrm{IW} -2$ and $f$O$_2 = \Delta \mathrm{IW}-4$ at tidal power densities $= 10^{-10}$ \si{W.kg^{-1}} are linearly extrapolated from other data points in figure (see data reduction code in Section \ref{sec:Data_avail}), as these cases solidified before tidal heating $>0.1$ \si{W.m^{-2}}. The purple lines shows the tidal Q parameter evolution found by \citet{ZAHNLE201574}, with times in Myr after the Moon-forming impact. The red regions show the Evection Resonance (ER) \citep{Touma_1998}, Laplace Plane Transition (LPT) \citep{Tremaine_2009}, and Cassini State Transition (CST) \citep{1969AJ.....74..483P}.}
    \label{fig:tidal_Q}
\end{figure}

\noindent Tidal heating is less efficient at large melt fractions because the mantle has a low viscosity which does not dissipate large amounts of heat when exposed to a tidal potential. When tidal heating becomes locally active in a layer of the mantle, upward progression of the rheological front is slowed, and volatile outgassing halts. Starting from a fully molten state, decreasing melt fractions yield increasingly efficient tidal heat dissipation, producing larger upward heat fluxes, until the rate of thermal emission from the surface is balanced by internal heat production and the local melt fraction becomes stable. The maximum tidal power dissipation is thus maximized during the GRE states, for a given set of parameters. Figure \ref{fig:tidal_Q} shows the tidal $Q$ parameter (efficiency) plotted against lunar orbital semi-major axis.

Oxidizing cases ($f$O$_2 > \Delta \mathrm{IW} + 0$) dissipate tidal heat most efficiently when the atmosphere is CO$_2$-dominated (Figure \ref{fig:tidal_Q}). Subsequently, H$_2$O entering the atmosphere makes these cases dissipate tidal heat only marginally more efficient. The assumption of a fixed tidal $Q$-factor is then justified in these cases. In contrast, reducing cases ($f$O$_2 < \Delta \mathrm{IW} + 0$) exhibit widely varying tidal $Q$-factors, even though their atmospheric composition remains largely unchanged in our simulations. \citet{ZAHNLE201574} report a similar relationship while assuming a fixed \textit{oxidizing} atmosphere under varying tidal parameters. Our findings show that the strong dependence of tidal efficiency on atmospheric composition introduces a notable degeneracy between tidal efficiency at different redox states. This phenomenon is not captured by models that assume a fixed atmosphere. Consequently, the assumption of fixed tidal parameters is not universally valid, especially for planets with fixed atmospheric composition and pressure \citep[e.g.][]{Canup2020}. 

Based on Figure \ref{fig:tidal_Q} we hypothesize that an oxidizing Earth would spin-down faster than a reducing Earth briefly after the moon-forming impact, with rapid early lunar migration. At later stages the oxidizing Earth likely remained tidally heated longer (Figure~\ref{fig:timelines_VMR}), at this point in time lunar migration is most rapid for the reduced Earth. Cases with $f$O$_2 = \Delta \mathrm{IW} + 0$ likely dissipate a given energy budget over the longest timescales, given the highly inefficient tidal heating limited by the strongest atmospheric greenhouse effects. This is in-line with cooling timescale trends calculated by \citet{Nicholls_2025_MNRAS} in the absence of tidal heating. Overall, this implies that cases with $f$O$_2 = \Delta \mathrm{IW} + 0$ will produce the slowest overall lunar migration. If the oxidization state of the Hadean atmosphere evolved from reducing to mildly oxidizing, then lunar orbital recession is most suppressed.

\subsection{Global radiative equilibrium epochs}\label{sec:GREdisc}

\noindent Periods of global radiative equilibrium are physical scenarios where energy dissipation to space is balanced by irradiation and internal heat production; e.g. Type II planets \citep{Hamano2013} or permanent magma ocean scenarios \citep{nicholls_tides_2025, Nguyen_TheEffect_2026}. Here, we find that magma ocean regimes on an early-Earth subject to tidal heating may attain a quasi-steady-state of global energy balance whilst the mantle is partially molten (Figure \ref{fig:Rheo_tide_timeline}A). Cases at different $f\text{O}_2$ and $H_\text{tide}$ demonstrate a stabilizing relationship between the tidal heating of planetary interiors, their temperature-dependent rheological properties, and radiative cooling to space. These cases all show a self-regulating melt fraction which is roughly constant until the tidal power is dissipated. Other works have proposed such self-regulating feedbacks to be present in the interiors of rocky bodies: within the Earth \citep{ZAHNLE201574}, on Io \citep{https://doi.org/10.1029/2002JE001943,OJAKANGAS1986341}, and within exoplanetary interiors \citep{Henning_2009,nicholls_tides_2025,farhat2025tides}. \citet{nicholls_tides_2025} self-consistently coupled the physics of magma ocean and atmospheric evolution under a range of different atmospheric scenarios, identifying prolonged and persistent magma oceans within the L\,98-59 system arising from this feedback. When a planet reaches such as quasi-steady-state in a regime of inefficient atmospheric escape, there will be no further evolution of the system, and thus the planet may remain in this state until tides are shut off. This increases the chances of observing active magma ocean phases in extrasolar planetary systems \citep{Bonati19,Cesario24}.

The modeled quasi-stable epochs were obtained under the assumption of fixed tidal power densities. This assumption only holds if the Moon’s orbital semi-major axis remains constant. \citet{tidestimes} show that, under constant tidal parameters, the Moon's outward migration is fixed shortly before and after the Laplace Plane Transition \citep{Tremaine_2009}. During the Laplace Plane Transition instability, Earth's spin decreases and angular momentum is transferred to Earth's orbit \citep{Cuk2016}. This idea is supported by the frequent occurrence of lunar rock ages around $4.35$\,Ga and a coincident spike in zircon ages \citep{Nimmo2024, doi:10.1126/sciadv.adn9871} which suggest a remelting event within the Moon linked to its orbital and tidal evolution. Moreover, constraints on lunar magma ocean solidification suggest that the lunar mantle underwent re-melting as recently as 2\,Ga \citep{nimmo2025surfaces, Byrne2020}, indicative of slow orbital recession rates. As shown in Figure~\ref{fig:tidal_Q}, such GRE epochs are most likely to occur for tidal power densities of $\sim 10^{-8}$–$10^{-9}$ [\si{W.kg^{-1}}] during the Laplace Plane Transition, and $\sim 10^{-9}$–$10^{-10}$ [\si{W.kg^{-1}}] during the Cassini State Transition. This scenario requires an initially higher Earth–Moon system angular momentum than its present value, which is favored by recent Moon-formation models. In particular, high--angular momentum cases may help explain the compositional similarity between Earth and Moon \citep{https://doi.org/10.1002/2016JE005239,Lock2018,Canup2020}. 

Inefficient tidal heating suppresses lunar orbital recession rates, making it more likely to be captured into one of these resonance states with the Sun \citep{Canup2020}. As such, we find that an Earth-like planet with $f$O$_2 = \Delta \mathrm{IW} + 0$ (and thus relatively inefficient tidal heat dissipation) is most likely to enter into a quasi-steady-state of global radiative equilibrium at some point during its evolution.

\citet{KORENAGA2025116743} propose that a tidally-supported stable configuration may occur when the lunar orbital semi-major axis has a value between 5 and 10 Earth radii. Our results further support this notion, by showing that for orbital separations $\sim 10 R_\mathTerra$ strong tidal heating (i.e. $H_\text{tide}$ $\approx 10^{-7}$ \si{W.kg^{-1}}) occurs (see Figure \ref{fig:tidal_Q}), which is able to maintain global melt fractions close to the critical melt fraction (Figure \ref{fig:melt_spread}).

In addition to this, more reducing cases maintain early GRE periods at greater Earth--Moon separations compared to oxidizing cases (Figure \ref{fig:melt_spread}) due to the relatively stronger greenhouse effects in reducing cases while the mantle melt fraction is large. However, the Moon likely only spent $\sim3\%$ of its lifetime at orbital separations $< 30 R_\mathTerra$ \citep{farhat2022resonant}, limiting the overall time interval of the GRE epochs.

\subsection{Prebiotic environment of the Hadean}

\subsubsection{Timescales of global radiative equilibrium}

\noindent Simulations performed with varying tidal power density show that oxidizing cases accommodate both short- and long-lasting GRE epochs which occur at a range of times following model initialisation. In comparison, reducing atmospheres accommodate short-lasting GRE epochs at tidal power densities $\geq 10^{-9}$ (topmost cases in Figure \ref{fig:timelines_VMR}A.)

The cases for which $f$O$_2 = \Delta \mathrm{IW} - 2$ take the longest to reach GRE. These cases reach the Laplace Plane Transition at around the same time found by \citet{abe2001tidal} and \citet{farhat2022resonant}, as well as \citet{KORENAGA2023115564} for their simulations with $Q_\mathTerra/k_{2 \ \mathTerra} \approx 10^2 - 10^3$. The cases for which $f$O$_2 = \Delta \mathrm{IW} + 0$ sustain tidal heating for the longest periods given the inefficient tidal heat dissipation (Figure \ref{fig:tidal_Q}). 

Late and long-lasting GRE epoch decrease the potential productivity of reduced meteoritic bombardement \citep{, Citron_2022,Wogan_2023}, as any impactor would fall onto a magma ocean surface, as has been demonstrated for M-dwarf exoplanets \citep{Lichtenberg2022ApJL}. If late bombardment did not substantially re-heat the atmosphere \citep{Citron_2022}, a later onset of GRE would reduce the gap between the GRE epoch and the earliest biosignatures. Long-lasting GRE periods could in principle support greater production of nitriles by extending the window for prebiotic chemistry. However, during the time spend in these states the surface is hot, and therefore not hospitable. 

Based on Hf-W dating, Rb-Sr dating, U-Pb dating, and dynamic simulations, the Moon-forming impact is expected to have occurred between 4.52 and 4.42\,Ga \citep{Touboul2007, doi:10.1098/rsta.2008.0209, doi:10.1126/sciadv.1602365, 2014Natur.508...84J}. Our results suggest that the Earth's mantle solidified at latest within 500\,Myr after the Moon-forming impact, similar to \citet{doi:10.1098/rsta.2015.0394}, which places the habitability boundary at 4.02-3.92\,Ga. \citet{Sole2025101126} recently found crystallization ages of $\approx$ 4.2 $\pm$ 0.1 Ga from oceanic rocks from the Nuvvuagittuq Greenstone Belt in northeast Canada. This age aligns well with our upper-limit estimates for magma ocean crystallization resulting from quasi-steady-states at intermediate redox states. These timelines are further in agreement with the possible earliest robust biosignatures imposed by observation of isotopic carbon signatures in rocks of sedimentary origin through tracer $\delta^{13}$C at 3.7\,Ga \citep{doi:10.1126/science.283.5402.674, ohtomo2013}, as well as stromatolites in metacarbonate rocks in the Isua Greenstone Belt dated to 3.7\,Ga \citep{Nutman2016}. In both cases this leaves $\sim 200$\,Myr to form oceans and for life to arise. A wet magma ocean, like simulated here, would form oceans shortly after solidification \citep{Miyazaki2022, https://doi.org/10.1002/jgre.20068}, while more conservative estimates range up to a few hundred Myr after accretion \citep{harrison2009hadean}. Prebiotic synthesis timescales are generally expected to be much shorter \citep{doi:10.1089/ast.2020.2335, Paschek_2025}, however these processes are highly sensitive to environmental thermo-chemical conditions \citep{Morasch2019NatCh,Ianeselli2022,Ianeselli2023NatRP}. The overall agreement between modeled timescales and observational boundaries underscores the validity of the employed modeling framework. As such, simulations of this kind may help constrain habitability periods elsewhere, such as on Venus, Mars, and exoplanets \citep{Krissansen-Totton_2021, salvador2023magma, Lichtenberg2025TOG3}. 

Genetic timing analysis of the Last Universal Common Ancestor (LUCA) of prokaryotic life suggests that LUCA lived between $4.09$ and $4.33$\,Ga \citep{Moody2024}. Only a subset of our simulations that assume tidal power densities $\geq 10^{-9}$ [\si{W.kg^{-1}}] are compatible with this time boundary. These cases dissipate energy faster (see Figure \ref{fig:sens_H_atm}) and deplete the energy budget quicker, thus solidifying sooner. 

\subsubsection{Water Distribution}

\noindent A wet mantle has been postulated as a key ingredient for a fast onset of mobile lid tectonics on the early Earth. \citet{Miyazaki2022} suggest that when the mantle is wet and dominated by Mg-rich pyroxenites, CO$_2$ removal from the atmosphere could be completed in 160\,Myr, potentially yielding habitable conditions for cases $f$O$_2 > \Delta \mathrm{IW} +0$ \citep[c.f.,][]{Lourenco2020GGG}. Our results show that tidal heating could have maintained a wet mantle for some time through the preferential dissolution of H$_2$O into the planet's interior. The wet mantle eventually produces an oceanic crust rich in olivine \citep{Miyazaki2022}, which promotes serpentinization, repartitioning CO$_2$ into the mantle \citep{Paschek_2025}. \citet{modirrousta2025efficacy} show that efficient magma ocean mixing \citep{ikoma2006constraints, young2023earth} can only be sustained at magma temperatures above $\sim 10,000$~K. Because our model does not produce such extreme temperatures through tidal heating, chemical exchange with the atmosphere may have been limited. This implies that the late release of water may require alternative interaction pathways. Despite this, future studies into ocean formation on an oxidizing Hadean Earth may benefit from the late release of water that we have demonstrated here. Especially, these studies may benefit from prolonged water retention and higher surface temperatures, due to tidal heating, allowing for greater CO$_2$ repartitioning. 

Our simulations highlight the physical relationship between melt fraction and H$_2$O interior-atmosphere partitioning. During tidally sustained global radiative equilibrium epochs, a smaller H$_2$O volume mixing ratio could allow stellar UV radiation to penetrate deeper into the atmosphere. In particular, UV photons can efficiently photolyze N$_2$, initiating reactions that have been shown to lead to efficient HCN formation in combination with CH$_4$ photolysis products \citep{TIAN2011417,1986JGR....91.2819Z,Ranjan2017}. Lower atmospheric H$_2$O concentrations can thus enhance N$_2$ fixation rates \citep{2016NatGe...9..452A,RIMMER2019124}, supporting the formation of important prebiotic feedstock species like HCN \citep{Urey1952,Patel2015,https://doi.org/10.1002/anie.201506585}. 

\begin{figure}[htbp]
    \centering
    \includegraphics{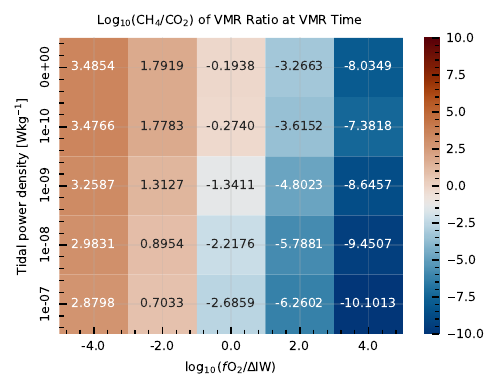}
    \vspace{-0.6cm}
    \caption{CH$_4$/CO$_2$ ratios across redox states in modeled atmospheres, showing a clear trend from solidified reducing (top left) to mostly molten oxidizing (bottom right) cases. High nitrile production is expected when CH$_4$/CO$_2 \sim 0.1$, i.e., $\sim$-1 on the logarithmic axis above.}
    \label{fig:ATM_best_ones}
\end{figure}

\subsubsection{HCN formation}

\noindent Detailed modeling of photochemical reaction pathways that produce hydrogen cyanide (HCN) generally suggests optimal yields when the CH$_4$/CO$_2$ molar ratio in the atmosphere is $\sim 0.1$ \citep{doi:10.1126/science.1183260,TIAN2011417,Wogan_2023}. Figure~\ref{fig:ATM_best_ones} illustrates how this ratio varies with modelled mantle redox state in our simulations, highlighting that the most optimal conditions for HCN formation occur between $\Delta$IW$-2 < f\mathrm{O}_2 < \Delta$IW$+0$. The mid-Hadean Earth potentially already possessed an oxidized mantle at $\sim 4.35$\,Ga  \citep{Trail2011}. At the time of solidification in our models, the $\Delta$IW$-2 < f\mathrm{O}_2 < \Delta$IW$+0$ range is mostly compatible with these constraints, making these intermediate redox states viable candidates for characterizing the early Earth's atmosphere and supporting prebiotic chemistry. A wider range of states could potentially be compatible if substantial amounts of hydrogen are lost through fractionating escape processes (\citet{doi:10.1126/science.1061976}; \citet{ZAHNLE201326}).

Once formed, HCN is subject to both photochemical destruction and settling to the surface. The dominant loss pathway is UV photolysis, \mbox{HCN + h$\nu \rightarrow$ CN + H}. According to \citet{TIAN2011417}, HCN production occurs primarily above 65\,km altitude, while its photolysis also becomes effective above this region, so efficient HCN stockpiling requires that it accumulates in the deeper atmosphere. An atmosphere with less H$_2$O would allow deeper UV penetration, potentially lowering the altitude of HCN formation and thereby increasing the risk of photolysis before it can reach the surface.

However, photochemical reaction rates under early Earth conditions remain largely extrapolated \citep[e.g.][]{white_rimmer_2024}. Consequently, we are currently unable to reliably model the exact chemical evolution of these early atmospheres. To advance our understanding of HCN formation in  Earth's early atmosphere, future studies should develop atmospheric chemistry at higher temperatures and pressures.

\subsection{Limitations}

\noindent Here, we have adopted a simplified parametrization of tidal heating in the planetary interior in order to remain agnostic the specific rheological models. This choice likely quantitatively changes our obtained tidal heating rates. For example, \citet{farhat2025tides} have recently expanded the established tidal heating formalism for magma ocean regimes by adding liquid-phase viscoelastic responses to tidal forcing. Hence, our formalism likely underestimates tidal dissipation in the fluid part of the mantle; the tidal efficiencies shown in Figure~\ref{eq:tidal_Q} may be underestimated. Notably, our derivation of the lunar orbital semi-major axis introduces substantial uncertainties in the specific horizontal placement of data points in Figure \ref{fig:tidal_Q}. Future works should improve on this through a coupled thermo-chemical-dynamic model of lunar orbital evolution alongside the Earth's interior and atmosphere. The application of a more realistic tidal heating calculation would yield a substantially reduced tidal heating rate within the near-solidified mantles considered in this work \citep{farhat2025tides}. This would likely cause our cases with tidal power densities $< 10^{-9}$ \si{W.kg^{-1}} to solidify sooner, and possibly also be compatible with the extrapolated time boundary imposed by LUCA \citep{Moody2024}. Ultimately, imposing a tidal shutdown condition based on the time in the current model, complementary to the energy budget, would be useful especially for cases with $H_\text{tide} < 10^{-9}$ \si{W.kg^{-1}}, allowing for earlier solidification and better alignment with previously discussed biosignature boundaries. Furthermore, we fix mantle $f$O$_2$ during the each simulation and explore a wide range from highly reduced to modern Earth-like. $f$O$_2$ should evolve in time with progressing core formation, hydrogen escape to space, and ongoing iron disproportionation \citep{Hirschmann2023EPSL,Schaefer2024}. In light of the estimated timescales for this \citep{Zahnle2020plas}, our oxidized cases can be regarded as the ones closest to geochemical expectations.

\section{Conclusion}\label{sec:conclusion}

\noindent We have explored the impact of tidal heating on the thermal and atmospheric evolution of the early Earth using the \texttt{PROTEUS} planetary evolution framework. Our results show that tidal heating could significantly influence the evolution of the early Earth by supporting periods of quasi-steady-state epochs with large outgoing radiation fluxes. By investigating a wide range of potential mantle redox states, we studied the interaction between tidal heating and atmospheric composition. Our findings are informative for hypotheses on the onset of habitable conditions and the origin of life on the early Earth. Our key results are:
\begin{itemize}
    \item Tidal power densities between $10^{-10}$ and $10^{-7}$~\si{W.kg^{-1}} significantly extend mantle solidification timescales from $\sim$30~Myr to $\sim$500~Myr.
    \item Weaker tidal scenarios enhance production of atmospheric H$_2$S and NH$_3$ and deplete CO, suggesting a potential pathway to observe the signature of disequilibrium processes in magma ocean atmospheres.
    \item Atmospheric heat loss to space due tidal heating varies by an order of magnitude across redox states, reaching up to 200~\si{W.m^{-2}}, depending on regulation by the atmospheric composition.
    \item Substantial H\textsubscript{2}O outgassing occurs near the end of solidification, enabling weaker tides to sustain larger melt fractions under more oxidizing conditions, whereas the phase state of reducing mantles are less sensitive to the strength of tidal heating.
    \item Periods of global radiative equilibrium emerge across a wide range of modeled tidally-heated scenarios. These quasi-steady-state epochs represent thermally stable configurations which can persist from 1.5 to 317~Myr, slowing planetary cooling.
    \item While at global radiative equilibrium, atmospheric CH\textsubscript{4}/CO\textsubscript{2} ratios near 0.1 were observed in our models for surface oxygen fugacities near the iron-w\"ustite buffer. Such conditions may enable the temporary accumulation of nitriles.
\end{itemize}

\noindent These findings highlight that the feedback between tidal heating and atmospheric composition may have played a crucial role in the Earth's transition to a habitable environment. While this study underscores the importance of tidal heating from an atmospheric perspective, future work should couple interior–atmosphere dynamics to the orbital evolution of the Earth–Moon system to further test the viability of this alternative pathway to habitability.

\section{Data Availability}\label{sec:Data_avail}
\noindent \texttt{PROTEUS}\footnote{\url{https://github.com/FormingWorlds/PROTEUS}}, \texttt{SPIDER}\footnote{\url{https://github.com/FormingWorlds/spider}}, and \texttt{AGNI}\footnote{\url{https://github.com/nichollsh/AGNI}} are open source software available on GitHub. The data reduction code underlying this article is available on the \texttt{tidal\_shutdown} branch of the \texttt{PROTEUS} Github repository, at \url{https://github.com/FormingWorlds/PROTEUS/tree/tidal_shutdown}. Specifically the \texttt{Data\_reduction\_code.ipynb} notebook is located in the \texttt{PROTEUS/tools/notebooks} folder.

\begin{acknowledgments}
\noindent T.L. thanks Simon Lock for enlightening discussions on lunar orbital recession. This project was supported by the Branco Weiss Foundation, the Netherlands eScience Center (PROTEUS project, NLESC.\-OEC.\-2023.\-017), the Alfred P. Sloan Foundation (AEThER project, G202114194), NASA’s Nexus for Exoplanet System Science research coordination network (Alien Earths project, 80NSSC21K0593), and the NWO NWA-ORC PRELIFE Consortium (PRELIFE project, NWA.1630.23.013). H.N. acknowledges support from STFC grant UKRI1184.
\end{acknowledgments}

\begin{contribution}
\textbf{MD:} Conceptualization, Methodology, Software, Validation, Formal analysis, Investigation, Writing - Original Draft, Writing - Review \& Editing, Visualization.
\textbf{HN:} Software, Methodology, Writing - Review \& Editing, Supervision.
\textbf{TL:} Conceptualization, Methodology, Validation, Formal analysis, Investigation, Writing - Review \& Editing, Software, Supervision.
\end{contribution}

\bibliography{sample701}{}
\bibliographystyle{aasjournalv7}

\end{document}